\documentclass{article}%
\usepackage{amsmath}
\usepackage{amsfonts}
\usepackage{amssymb}
\usepackage{graphicx}%
\begin{document}

\begin{center}

\Large{\bf Generating Solutions to the Einstein Field Equations}\\

\vspace{1.cm}

\large{I. G. Contopoulos$^1$, F. P. Esposito$^2$, K. Kleidis$^3$,}\\
\large{D. B. Papadopoulos$^4$, and L. Witten$^5$}\\

\vspace{.5cm}

{\small $^1$Research Center for Astronomy and Applied Mathematics,}\\
{\small Academy of Athens, 115 27 Athens, Greece}\\
{\small $^2$Department of Physics, University of Cincinnati, 45269 Ohio, USA}\\
{\small $^3$Department of Mechanical Engineering, Technological Education Institute of Central Macedonia, 621 24 Serres, Greece}\\
{\small $^4$Department of Physics, Aristotle University of Thessaloniki,}\\
{\small 541 24 Thessaloniki, Greece} \\
{\small $^5$Department of Physics, University of Florida, 32611-8440 Florida, USA}

\end{center}

\bigskip

\begin{abstract}

Exact solutions to the Einstein field equations may be generated from already existing ones (seed solutions), that admit at least one Killing vector. In this framework, a space of potentials is introduced. By the use of symmetries in this space, the set of potentials associated to a known solution are transformed into a new set, either by continuous transformations or by discrete transformations. In view of this method, and upon consideration of continuous transformations, we arrive at some exact, stationary axisymmetric solutions to the Einstein field equations in vacuum, that may be of geometrical or/and physical interest.

\end{abstract}

\section{Introduction}

Originating from an exact solution to the equations of General Relativity (GR) in vacuum, that admits a congruence of Killing vectors, one may generate other solutions~\cite{1}. The Killing vectors of the original solution are assumed to have a non-zero norm and a twist potential, both of which are accordingly transformed to construct the Killing vectors of the resulting (new) solution.

In the present article, we give several examples on how to use symmetries in the (so called) potential space, to generate stationary axisymmetric solutions to the Einstein field equations (see, e.g.,~\cite{2}). To do so, we follow the method developed by Ehlers and others from the early 60s till the late 70s (see, e.g., Chapter 34.1 - 34.3 of~\cite{1}). In view of this method, any spacetime metric can be expressed in two equivalent ways; either in the form of the line element \begin{equation} ds^2 = \sigma \left ( dt + A_i dx^i \right)^2 + h_{ij} dx^i dx^j \: , \end{equation} (in the system of units where both Newton's constant and the velocity of light equal to unity) or as the symmetric tensor \begin{equation} g_{\mu \nu} = h_{\mu \nu} + \frac{1}{\sigma} k_{\mu} k_{\nu} \: . \end{equation} In Eqs. (1) and (2), Greek indices refer to the four-dimensional spacetime, Latin indices refer to the three-dimensional slice that is perpendicular to a particular Killing vector, $k_{\mu}$, and $h_{ij}$ is the three-metric of such a slice. Moreover, in Eq. (1), the three-vector $A_i$ is defined by \begin{equation} k_i = \sigma A_i \: , \end{equation} where $\sigma$ is the negative norm of the Killing vector $k_{\mu}$, i.e., \begin{equation} k_{\mu}k^{\mu} = - \sigma \: . \end{equation} 

There is also a twist potential, $\omega$, associated with $k_{\mu}$, defined by \begin{equation} \omega_{\: , \: \mu} = \eta_{\mu \nu \rho \sigma} k^{\nu} \partial^{\rho} k^{\sigma} \: . \end{equation} In Eq. (5), the comma denotes partial derivative, $\eta_{\mu \nu \rho \sigma} = \sqrt{\vert \: g \: \vert} \: \varepsilon_{\mu \nu \rho \sigma}$, where $g$ is the determinant of the metric tensor (2) and $\varepsilon_{\mu \nu \rho \sigma}$ is the completely antisymmetric symbol of four indices, with $\varepsilon_{0 1 2 3} = 1$. 

Upon consideration of $\sigma$ and $\omega$, a complex potential, the well-known Ernst potential, $\epsilon$, and a conformal change of the three-metric are defined, as follows \begin{eqnarray} && \epsilon = \sigma + i \omega \: , \\ && \widetilde{h}_{ij} = \sigma h_{ij} \: . \end{eqnarray} Using both of these quantities, the new solution is accordingly generated by \begin{eqnarray} \epsilon^{\prime} & = & \frac{\epsilon}{1 + i \alpha \epsilon} \\ \mbox{~and~}~~ \widetilde{h}_{ij}^{\prime} & = & \widetilde{h}_{ij} \end{eqnarray} (see, e.g., Eq. 34.11 of~\cite{1}). In view of Eq. (8), the new Ernst potential is obtained by means of a Kinnersley IV transformation~\cite{3}, that introduces the parameter $\alpha$, which throughout our analysis is considered to be real. This transformation has (accordingly) been generalized to include an infinite number of new parameters, $\alpha_n$ $(n = 1, 2, 3,...)$, in the context of the Hoenselaers-Kinnersley-Xanthopoulos transformation~\cite{2}. 

In Eqs. (8) and (9), the original (seed) metric is denoted without a prime, while the generated (new) metric is indicated with a prime. Assuming that the twist potential of the original solution is zero, the norm of the new Killing vector, $\sigma^{\prime}$, the new twist potential, $\omega^{\prime}$, and the new three-metric, $h_{ij}^{\prime}$, are given by \begin{eqnarray} \sigma^{\prime} & = & \frac{\sigma}{1 + \alpha^2 \sigma^2} \: , \\ \omega^{\prime} & = & - \frac{\alpha \sigma^2}{1 + \alpha^2 \sigma^2} \\ \mbox{~and~}~~ h_{ij}^{\prime} & = & \frac{\sigma}{\sigma^{\prime}} \: h_{ij} \: , \end{eqnarray} respectively. Now, the components of the Killing vector, $k_{\mu}^{\prime}$, corresponding to the generated metric, can be obtained by solving the definition equation of the new twist potential, $\omega^{\prime}$ (cf. Eq. 5).

In what follows, we shall use the aforementioned technique, to generate twisting, stationary axisymmetric solutions of GR, that originate from any static axisymmetric solution given in Weyl (cylindrical) coordinates. As we shall demonstrate, either the spacelike, $\frac{\partial}{\partial \phi}$, or the timelike, $\frac{\partial}{\partial t}$, Killing vector can be used, to generate the new twist potential. More specifically:

In Sect. 2, we give the particulars in detail, using the spacelike Killing vector. We then demonstrate the method, by applying it to the Schwarzschild solution, thus resulting in a new solution to the Einstein field equations. In Sect. 3, the method is applied to any static axisymmetric solution using the timelike Killing vector. Accordingly, another exact solution is generated, originating, once again, from the Schwarzschild one. This solution turns out to be the Taub-NUT metric~\cite{4},~\cite{5}. Finally, in Sect. 4, we apply the method of generating solutions (upon the action of the timelike Killing vector) to the $\gamma$-metric~\cite{6},~\cite{7} (also called the Zipoy-Voorhees solution).

\section{Solutions generated from static axisymmetric ones, using the spacelike Killing vector}

In Weyl coordinates, static axisymmetric solutions of GR are described by~\cite{8} \begin{equation} ds^2 = e^{2 \lambda} dt^2 - e^{- 2 \lambda} \left [ e^{2 \mu} \left ( d \rho^2 + dz^2 \right ) + \rho^2 d \phi^2 \right ] \: , \end{equation} where \begin{eqnarray} && \lambda_{\: , \: \rho \rho} + \frac{\lambda_{\: , \: \rho}}{\rho} + \lambda_{\: , z z} = 0 \: , \nonumber \\ && \mu_{\: , \: \rho} = \rho \left ( \lambda_{\: , \: \rho}^2 - \lambda_{\: , \: z}^2 \right ) \: , \\ && \mu_{\: , \: z} = 2 \rho \: \lambda_{\: , \: \rho} \: \lambda_{\: , \: z} \: . \nonumber \end{eqnarray} The metric given by Eq. (13) admits two Killing vectors, one spacelike, $\partial / \partial \phi$, and one timelike, $\partial / \partial t$. Hence, originating from (13), we expect that we can generate new solutions to the Einstein field equations in vacuum, using either of these two Killing vectors.

In the axisymmetric solution given by Eq. (13), the norm of the spacelike Killing vector, $k_{\phi} = \partial_{\phi}$, is $\sigma = \rho^2 \exp(- 2 \lambda)$, so that $k_{\phi}k^{\phi} = - \sigma < 0$ (as it should), and the twist potential vanishes $(\omega = 0)$; hence, in this case, the Ernst potential of the seed solution (cf. Eq. 6) reads \begin{equation} \epsilon = \sigma = \rho^2 e^{-2 \lambda} \: . \end{equation} In order to generate a solution that originates from (13), we first use Eq. (12), to determine the new three-metric, $h_{ij}^{\prime}$, as \begin{equation} 
h^{\prime}_{ij} = \frac{\sigma}{\sigma^{\prime}} \left ( \begin{array}
[c]{ccc}
e^{2 \lambda} & 0                      & 0                        \\
0             & - e^{2 \mu -2 \lambda} & 0                        \\
0             & 0                      & - e^{2 \mu - 2 \lambda}
\end{array} \right ) \: , \end{equation} with determinant $h^{\prime} = \det \vert \vert h_{ij}^{\prime} \vert \vert = \left ( \frac{\sigma}{\sigma^{ \prime}} \right )^3  e^{4 \mu - 2 \lambda}$, where, in view of Eq. (10), \begin{equation} \frac{\sigma}{\sigma^{\prime}} = 1 + \alpha^2 \sigma^2 \: . \end{equation} 

On the other hand, the new twist potential is defined by \begin{equation} {\omega^{\prime}}^{\: , \: \mu} = \eta^{\mu \beta \gamma \delta} k_{\beta}^{\prime} \partial_{\gamma} k_{\delta}^{\prime} \: , \end{equation} where $k_{\beta}^{\prime}$ is the new Killing vector and \begin{equation} {\omega^{\prime}}^{\: , \: \mu} = g^{\mu \nu} {\omega^{\prime}}_{, \: \mu} = \left ( 0, \: {h^{\prime}}^{\rho \rho} {\omega^{ \prime}}_{\: , \: \rho} , \: {h^{\prime}}^{z z} {\omega^{\prime}}_{\: , \: z} \: , 0 \right ) \: . \end{equation} 

Now, all we have to do, is to determine the components of the vector $A_i ~(i = t, \: \rho, \: z)$, and, through those, the exact form of $k_i^{\prime} = \sigma^{\prime} A_i$, as well. Since we are interested in axisymmetric solutions of GR, in what follows, we assume that all functions depend solely on $\rho$ and $z$. In this context, it can be shown that the components $A_{\rho} = constant = A_z$ and may be chosen to vanish. Accordingly, \begin{equation} k_{\beta}^{\prime} = (\underbrace{\sigma^{ \prime} A_t , 0 , 0}_{k_i^{\prime}} , \sigma^{\prime}) \: . \end{equation}

Upon consideration of Eq. (20), Eq. (18) is written in the form \begin{equation} {\omega^{\prime}}^{\: , \: \rho} = \eta^{\rho \beta \gamma \delta} k_{\beta}^{\prime} \partial_{\gamma} k_{\delta}^{\prime} = \eta^{t \rho z \phi} \left [ k_{\phi}^{\prime} \partial_z  k_t^{\prime} - k_t^{\prime} \partial_z k_{\phi}^{\prime} \right ] \: , \end{equation} where \begin{equation} - \eta^{\rho t z \phi} = \eta^{t \rho z \phi} \: ,~~\eta^{\rho \phi z t} = \eta^{t \rho z \phi}~ \mbox{~and~}~~ \eta^{t \rho z \phi} \eta_{t \rho z \phi} = - 1 \: . \end{equation} Accordingly, we multiply both sides of Eq. (21) by $\eta_{t \rho z \phi}$. The left hand side yields \begin{equation} \eta_{t \rho z \phi} {h^{\prime}}^{\rho \rho} {\omega^{\prime}}_{\: , \: \rho} = \left( \vert \sigma^{\prime} h^{\prime} \vert \right )^{1/2} {h^{\prime}}^{\rho \rho} {\omega^{\prime}}_{\: , \: \rho} \: , \end{equation} while the right hand side results in \begin{equation} \eta_{t \rho z \phi} \eta^{t \rho z \phi} \left [ {k^{\prime}}_{\phi} \partial_z {k^{\prime}}_t - {k^{\prime}}_t \partial_z {k^{\prime}}_{\phi} \right ] = - \left (\sigma^{\prime} \right )^2  \partial_z A_t \: , \end{equation} where $\eta_{t \rho z \phi} = \sqrt{ \vert \sigma^{\prime} h^{\prime} \vert } \: \varepsilon_{t \rho z \phi}$ and $\varepsilon_{t \rho z \phi} = 1$. Consequently, upon consideration of Eqs. (23) and (24), Eq. (21) leads to \begin{equation} \left ( \vert \sigma^{\prime} h^{\prime} \vert \right )^{1/2} {h^{\prime}}^{ \rho \rho} {\omega^{ \prime}}_{\: , \: \rho} = - \left ( \sigma^{\prime} \right )^2 \partial_z A_t \end{equation} and \begin{equation} \left ( \vert \sigma^{\prime} h^{\prime} \vert \right)^{1/2} {h^{\prime}}^{z z} {\omega^{\prime}}_{\: , \: z} = \left ( \sigma^{\prime} \right )^2 \partial_{\rho} A_t \: . \end{equation} Further manipulation of Eq. (25) yields \begin{eqnarray} \left ( \vert \sigma^{\prime} h^{\prime} \vert \right )^{1/2} {h^{\prime}}^{\rho \rho} {\omega^{\prime}}_{\: ,\: \rho} & = & \left [ - \frac{\sigma^{\prime}}{ \sigma} e^{2 \lambda - 2 \mu} \right ] \left ( \vert \sigma^{\prime} h^{\prime} \vert \right )^{1/2} {\omega^{\prime}}_{\: , \: \rho} \nonumber \\ & = & - \rho {\omega^{ \prime}}_{\: , \: \rho} \end{eqnarray} and, similarly, from Eq. (26) we obtain \begin{eqnarray} \left ( \vert \sigma^{\prime} h^{\prime} \vert \right )^{1/2} {h^{\prime}}^{z z} {\omega^{\prime}}_{\: , \: z} & = & \left [ - \frac{\sigma^{\prime}}{\sigma} e^{2 \lambda - 2 \mu} \right ] \left ( \vert \sigma^{\prime}  h^{ \prime} \vert \right )^{1/2} {\omega^{\prime}}_{\: ,\: z} \nonumber \\ & = & - \rho {\omega^{\prime}}_{\: , \: z} \: . \end{eqnarray} However, in view of Eq. (11), we have \begin{eqnarray} && \frac{\partial \omega^{\prime}}{ \partial \rho} = \frac{\partial \omega^{\prime}}{\partial \sigma} \frac{\partial \sigma}{\partial \rho} = - \frac{4 \alpha \sigma^2}{\left ( 1 + \alpha^2 \sigma^2 \right )^2} \left ( \frac{1}{\rho} - \lambda_{\: , \: \rho} \right ) \\ && \frac{\partial \omega^{\prime}}{\partial z} = \frac{\partial \omega^{\prime}}{\partial \sigma} \frac{\partial \sigma}{\partial z} = \frac{4 \alpha \sigma^2}{\left ( 1 + \alpha^2 \sigma^2 \right )^2} \lambda_{\: , \: z} \: . \end{eqnarray} Inserting Eqs. (29) and (30) into Eqs. (27) and (28), respectively, and taking into account (also) Eq. (15), we obtain \begin{eqnarray} 4 \alpha \left ( 1 - \rho \frac{\partial \lambda}{\partial \rho} \right ) & = & - \frac{\partial A_t}{ \partial z} \: , \\ 4 \alpha \rho \frac{\partial \lambda}{\partial z} & = & - \frac{\partial A_t}{\partial \rho} \: . \end{eqnarray} We observe that, as far as the system of Eqs. (31) and (32) is concerned, the integrability condition is satisfied, simply by reference to the definition equation of $\lambda$ (the first one of Eqs. 14).

Summarizing, the solution generated from the seed metric (13) by the action of the $\phi$-Killing vector, is given by \begin{eqnarray} ds^2 & = & \left ( 1 + \alpha^2 \rho^4 e^{- 4 \lambda} \right ) \left [ e^{2 \lambda} dt^2 - e^{2 \mu - 2 \lambda} \left ( d \rho^2 + dz^2 \right ) \right ] \nonumber \\ & - & \frac{\rho^2 e^{- 2 \lambda}}{1 + \alpha^2 \rho^4 e^{- 4 \lambda}} \left (d \phi + A_t dt \right )^2 \: , \end{eqnarray} together with the system of Eqs. (14), as well as Eqs. (31) and (32).

\subsection{New solutions generated from Schwarzschild}

We shall apply the aforementioned technique of generating solutions, using as seed metric the Schwarzschild solution. In cylindrical coordinates, the Schwarzschild solution admits the form (13) with \begin{eqnarray} e^{2 \lambda} & = & \frac{L - m}{L + m} \: , \nonumber \\ \mbox{~and~}~~ e^{2 \mu - 2 \lambda} & = & \frac{(L + m)^2}{l_{+} \: l_{-}} \: . \end{eqnarray} In Eqs. (34), $m$ is the mass of the central object that is responsible for the spherically symmetric gravitational field and \begin{equation} L = \frac{1}{2} \left ( l_{+} + l_{-} \right ) \: , \end{equation} where \begin{equation} l_{\pm}^2 = \rho^2 + (z \pm m)^2 \: . \end{equation} 

Upon consideration of Eqs. (34) - (36), we can now integrate Eqs. (31) and (32), to obtain \begin{equation} A_t = 2 \alpha \left [ \left ( l_{+} - l_{-} \right ) - 2z \right ] \: . \end{equation} Hence, in view of Eqs. (33) and (37), the new metric generated from Schwarzschild reads \begin{eqnarray} ds^2 & = & \frac{1}{(L-m)^2} \left [ (L-m)^2 + \alpha^2 \rho^4 (L+m)^2 \right ] \left [ \frac{L-m}{L+m} dt^2 - \frac{(L+m)^2}{l_{+} \: l_{-}} \left ( d \rho^2 + dz^2 \right ) \right ] \nonumber \\ & - & \frac{\rho^2 \left (L^2 - m^2 \right )}{(L - m)^2 + \alpha^2 \rho^4 (L+m)^2} \left \lbrace d \phi + 2 \alpha \left [ \left (l_{+} - l_{-} \right ) - 2z \right ] dt \right \rbrace^2 \: . \end{eqnarray} 

Furthermore, performing the coordinate transformation $(t, \rho, z, \phi) \Rightarrow (t, r, \theta, \phi)$, where \begin{equation} \rho = \sqrt{r^2 - 2 m r} \sin{\theta} \: ,~~z = (r-m) \cos{\theta} \: ,~~L = r-m \: ,~~l_{+} - l_{-} = 2m \cos{\theta} \: , \end{equation} we find that, the metric given by Eqs. (13) and (34) - (36) reduces to the Schwarzschild solution in spherical $(t, r, \theta, \phi)$ coordinates, \begin{equation} ds^2 = \left ( 1 - \frac{2m}{r} \right) d t^2 - \frac{d r^2}{1 - \frac{2m}{r}} - r^2 \left (d \theta^2 + \sin^2 \theta d \phi^2 \right ) \: , \end{equation} as it should; while, the metric given by Eq. (38) results in \begin{eqnarray} ds^2  & = & \left ( 1 + \alpha^2 r^4 \sin^4 \theta \right ) \left [ \left ( 1 - \frac{2m}{r} \right )  dt^2 - \left ( 1 - \frac{2m}{r} \right )^{-1} dr^2 - r^2 d \theta^2 \right ] \nonumber \\ & - & \frac{r^2 \sin^2 \theta}{\left ( 1 + \alpha^2 r^4 \sin^4 \theta \right ) } \left [ d \phi - 4 \alpha (r - 2m) \cos \theta dt \right ]^2 \: . \end{eqnarray} 

The Riemann-square invariant of the new solution (41), is given by \begin{eqnarray} I & = & {\cal R}_{\mu \nu \lambda \rho} {\cal R}^{\mu \nu \lambda \rho} \nonumber \\ & = & \frac{1}{r^6 \left [ 1 + \alpha^2 r^4 \sin^4 \theta \right]^6} \left \lbrace -12 \alpha^8 r^{16} \sin^{12} \theta \left [ 12 r r_s \sin^4 \theta \right . \right . \nonumber \\ & - & \left . 16 r^2 - 21 r_s^2 \sin^4 \theta + 24 r r_s \sin^2 \theta \right ] \\ & + & 48 \alpha^6 r^{12} \sin^8 \theta \left [ - 70 r_s^2 \sin^4 \theta - 60 r^2 + 27 r r_s \sin^4 \theta + 102 r r_s \sin^2 \theta \right ] \nonumber \\ & + & 120 \alpha^4 r^8 \sin^4 \theta \left [ 21 r_s^2 \sin^4 \theta - 52 r r_s \sin^2 \theta + 24 r^2 + 6 r r_s \sin^4 \theta \right ] \nonumber \\ & - & \left . 96 \alpha^2 r^5 \left [ - 18 r_s \sin^2 \theta + 15 r_s \sin^4 \theta + 4r \right ] + 12 r_s^2 \right \rbrace \: , \nonumber \end{eqnarray} where $r_s = 2m$. In the limit of vanishing $m$, the invariant given by Eq. (42) does not vanish, and, therefore, we have generated a new solution from vacuum. On the other hand, for $m \neq 0$ and $\alpha = 0$, we obtain \begin{equation} I = {\cal R}_{\mu \nu \lambda \rho} {\cal R}^{\mu \nu \lambda \rho} = \frac{12 r_s^2}{r^6} = I_{Schw} \: , \end{equation} i.e., we recover the Riemann-square invariant associated to Schwarzschild solution; a not unexpected result, since, every difference in the metric given by Eq. (41), as compared to the Schwarzschild one (Eq. 40), is due to the terms involving $\alpha$. For $\alpha \neq 0$, these changes result in a dramatic variance on the $z ( = r \sin \theta)$ dependence. Furthermore, there is an additional rotation experienced by the coordinate system, which changes its direction at $r = 2m$, as well as at $\theta = \pi / 2$.

\section{Solutions generated from static axisymmetric ones, using the timelike Killing vector}

As far as the metric (13) is concerned, the timelike Killing vector associated with it, $k_t = \partial_t$, has the norm $\sigma = - \exp (2 \lambda)$, so that $k_t k^t = - \sigma > 0$, as it should. In this case, since (once again) $\omega = 0$, the seed potential is given by \begin{equation} \epsilon = \sigma = - \exp (2 \lambda) \: . \end{equation} Now, according to Eq. (12), the new three-metric tensor is written in the form \begin{equation} h_{ij}^{\prime} = \left ( 1 + \alpha^2 e^{4 \lambda} \right ) \left ( \begin{array}
[c]{ccc}
- e^{2 \mu - 2 \lambda} & 0                     & 0 \\
0                     & - e^{2 \mu - 2 \lambda} & 0 \\
0                     & 0                     & - \rho^2 e^{- 2 \lambda}
\end{array} \right) \: , \end{equation} with determinant $h^{\prime} = \det \vert \vert h_{ij}^{\prime} \vert \vert = - \left ( 1 + \alpha^2 e^{4 \lambda} \right )^3 \rho^2 e^{4 \mu - 6 \lambda}$. Next, we have to determine the components of the new Killing vector, \begin{equation} k_{\beta}^{\prime} = \left ( \sigma^{ \prime}, \sigma^{ \prime} A_i \right ),~~i = \rho, z, \phi \: , \end{equation} i.e., we need to determine $A_i$. To do so, we note that, the twist potential corresponding to the new solution, $\omega^{\prime}$, as given by Eq. (18), yields \begin{equation} \eta_{\rho z \phi t} {\omega^{\prime}}^{\: , \: \rho} = k_z^{\prime} \left ( \partial_{\phi} k_t^{\prime} - \partial_t k_{\phi}^{\prime} \right) + k_{\phi}^{\prime} \left( \partial_t k_z^{\prime} - \partial_z k_t^{\prime} \right ) + k_t^{\prime} \left ( \partial_z k_{\phi}^{\prime} - \partial_{\phi} k_z^{\prime} \right ) \: . \end{equation} Once again, axial symmetry implies that all functions depend solely on $\rho$ and $z$ and, therefore, $A_{\rho} = 0 = A_z$. Accordingly, by virtue Eq. (46), Eq. (47) results in \begin{equation} \left ( \vert \sigma^{\prime} h^{\prime} \vert \right )^{1/2} {h^{\prime}}^{\rho \rho} \omega_{\: , \: \rho}^{\prime} = \left ( \sigma^{\prime} \right )^2 \partial_z A_{\phi} \: , \end{equation} and \begin{equation} \left ( \vert \sigma^{ \prime} h^{\prime} \vert \right )^{1/2} {h^{\prime}}^{z z} {\omega^{\prime}}_{\: , \: z} = - \left (\sigma^{\prime} \right )^2 \partial_{\rho} A_{\phi} \: . \end{equation} In this case, in view of Eq. (11), and taking into account Eq. (44), we have \begin{eqnarray} && \frac{\partial \omega^{\prime}}{\partial \rho} = \frac{\partial \omega^{\prime}}{\partial \sigma} \frac{\partial \sigma}{\partial \rho} = - \frac{4 \alpha \sigma^2}{\left ( 1 + \alpha^2 \sigma^2 \right )^2} \lambda_{\: , \: \rho} \: , \\ && \frac{\partial \omega^{\prime}}{\partial z} = \frac{\partial \omega^{\prime}}{\partial \sigma} \frac{\partial \sigma}{\partial z} = - \frac{4 \alpha \sigma^2}{\left ( 1 + \alpha^2 \sigma^2 \right )^2} \lambda_{\: , \: z} \: . \end{eqnarray} Now, inserting Eqs. (50) and (51) into Eqs. (48) and (49), respectively, we obtain \begin{eqnarray} 4 \alpha \rho \frac{\partial \lambda}{\partial \rho} & = & \frac{\partial A_{\phi}}{\partial z} \: , \\ 4 \alpha \rho \frac{\partial \lambda}{\partial z} & = & - \frac{\partial A_{\phi}}{\partial \rho} \: , \end{eqnarray} where, once again, integrability of Eqs. (52) and (53) is guaranteed by the definition equation of $\lambda$ (the first one of Eqs. 14).

Summarizing, the solution generated from the seed metric (13) by the action of the $t$-Killing vector, is written in the form \begin{equation} ds^2 = \frac{e^{2 \lambda}}{1 + \alpha^2 e^{4 \lambda}} \left( dt + A_{\phi} d\phi \right )^2 - \frac{1 + \alpha^2 e^{4 \lambda}}{e^{2 \lambda}} \left [ e^{2 \mu} \left ( d \rho^2 + dz^2 \right ) + \rho^2 d \phi^2 \right ] \: , \end{equation} together with the system of Eqs. (14), as well as Eqs. (52) and (53). Gautreau and Hoffman~\cite{9} have also shown how to generate new twisting solutions from a seed metric in Weyl coordinates, using another method, one that was pioneered by Papapetrou~\cite{10}. It is worth noting that, although originating from the same seed metrics, their method generates other solutions than those we find in the present article.

\subsection{New solutions generated from Schwarzschild}

Once again, we shall apply the method presented in Sect. 3, using as seed metric the Schwarzschild solution in cylindrical coordinates. By virtue of Eqs. (34) - (36), the system of Eqs. (52) and (53) is directly integrated, yielding \begin{equation} A_{\phi} = \pm 2 \alpha \left ( l_{+} - l_{-} \right ) \: . \end{equation} Accordingly, in view of Eqs. (54) and (55), an exact solution that is generated from the seed metric (13) by the action of the $t$-Killing vector arises, namely, \begin{eqnarray} ds^2 & = & \frac{e^{2 \lambda}}{1 + \alpha^2 e^{4 \lambda}} \left [ dt - 2 \alpha \left ( l_{+} - l_{-} \right ) d \varphi \right ]^2 \nonumber \\ & - & \frac{1 + \alpha^2 e^{4 \lambda}}{e^{2 \lambda}} \left [ e^{2 \mu} \left ( d \rho^2 + dz^2 \right ) + \rho^2 d \phi^2 \right ] \: . \end{eqnarray}

To further scrutinize this result, we first perform the transformation given by Eq. (39), in order to obtain the solution (56) in spherical coordinates. Accordingly, Eq. (55) yields \begin{equation} A_{\phi} = - 4 \alpha m \cos \theta \end{equation} and the metric (56) is written in the form \begin{eqnarray} ds^2 & = & \frac{r^2 - 2 m r}{r^2 + \alpha^2 \left ( r - 2 m \right )^2} \left ( dt - 4 \alpha  m \cos \theta d \phi \right )^2 \nonumber \\ & - & \frac{r^2 + \alpha^2 \left ( r - 2 m \right )^2}{r^2 - 2 m r} dr^2 - \left [ r^2 + \alpha^2 \left ( r - 2 m \right )^2 \right ] d \Omega^2 \: , \end{eqnarray} where $d \Omega^2 = r^2 \left ( d \theta^2 + \sin^2 \theta d \phi^2 \right )$. 

A question arises now, on whether we can recover Eq. (58), using as seed metric the Schwarzschild solution in spherical coordinates (40), which admits the timelike Killing vector $\xi_t = \partial_t$ with norm $\sigma = - \left ( 1 - \frac{2m}{r} \right )$. In this case, \begin{eqnarray} \omega & = & 0 \, \Rightarrow \, \epsilon = \sigma = - \left ( 1 - \frac{2m}{r} \right ) \\ \mbox{~and~}~~h_{ij} & = & - \left ( \begin{array}
[c]{ccc}
\frac{1}{1 - \frac{2m}{r}} & 0   & 0 \\
0                          & r^2 & 0 \\
0                          & 0   & r^2 \sin^2 \theta
\end{array} \right ) \: . \end{eqnarray} Now, the combination of Eqs. (12), (17) and (60) yields the three-metric of the new solution, as \begin{eqnarray} && h_{ij}^{\prime} = - \left [ 1 + \alpha^2 \left ( 1 - \frac{2m}{r} \right )^2 \right ] \: \left ( \begin{array}
[c]{ccc}
\frac{1}{1 - \frac{2m}{r}} & 0     & 0 \\
0                          & r^{2} & 0 \\
0                          & 0     & r^2 \sin^2 \theta
\end{array} \right ) \: . \end{eqnarray} Furthermore, upon consideration of Eq. (46), Eq. (18) results in the following system of equations \begin{eqnarray} && \eta_{t r \theta \phi} {h^{\prime}}^{r r} {\omega^{\prime}}_{\: , \: r} = k_t^{\prime} \partial_{\theta} k_{\phi}^{\prime} = \left ( \sigma^{\prime} \right )^2 \partial_{\theta} A_{\phi} \: , \\ && \eta_{t r \theta \phi} {h^{\prime}}^{\theta \theta} {\omega^{\prime}}_{\: , \: \theta} = k_{\phi}^{\prime} \partial k_t^{\prime} - k_t^{\prime} \partial_r k_{\phi}^{\prime} = - \left ( \sigma^{\prime} \right )^2 \partial_r A_{\phi} \: , \end{eqnarray} where, by virtue of Eq. (59), from Eq. (11) we obtain \begin{eqnarray} \frac{\partial \omega^{\prime}}{\partial r} & = & \frac{\partial \omega^{\prime}}{\partial \sigma}\frac{\partial \sigma}{\partial r} = \frac{4 \alpha m \sigma}{r^2 \left ( 1 + \alpha^2 \sigma^2 \right )^2} \\ \mbox{~and~}~~\frac{\partial \omega^{\prime}}{\partial \theta} & = & \frac{\partial \omega^{\prime}}{\partial \sigma} \frac{\partial \sigma}{\partial \theta} = 0 \: . \end{eqnarray} In view of Eqs. (64) and (65), Eqs. (62) and (63) yield \begin{eqnarray} - \sqrt{\vert \: \sigma \: \vert} \: \sin \theta \sqrt{1 - \frac{2m}{r}} \frac{ 4 \alpha m \sigma}{\left ( 1 + \alpha^2 \sigma^2 \right )^2} & = & \frac{\sigma^2}{\left ( 1 + \alpha^2 \sigma^2 \right )^2} \partial_{\theta} A_{\phi} \: , \nonumber \\ 0 & = & \partial_r A_{\phi} \end{eqnarray} so that, eventually, \begin{eqnarray} \partial_{\theta} A_{\phi} & = & 4 m \alpha \sin \theta \sqrt{ \frac{\vert \: \sigma \: \vert}{1 - \frac{2m}{r}}} \: , \nonumber \\ \partial_r A_{\phi} & = & 0 \: . \end{eqnarray} Upon consideration of Eq. (59), Eqs. (67) result in \begin{equation} A_{\phi} = - 4 \alpha m \cos \theta \end{equation} and, therefore, the metric generated from Schwarzschild, by the action of the $t$-Killing vector, is written in the form \begin{eqnarray} ds^2 & = & \frac{r^2 - 2 m r}{r^2 + \alpha^2 \left ( r - 2 m \right )^2} \left ( dt - 4 \alpha m \cos \theta d \phi \right )^2 \nonumber \\ & - & \frac{r^2 + \alpha^2 \left ( r - 2 m \right )^2}{r^2 - 2 m r} dr^2 - \left [ r^2 + \alpha^2 \left ( r - 2 m \right )^2 \right ] d \Omega^2 \: , \end{eqnarray} which, indeed, coincides with Eq. (58). Hence, as far as the Schwarzschild solution is concerned, and upon the action of the $t-$Killing vector, these two operations (coordinate transformations and the solutions generating mechanism) commute with each other.

The metric given by Eq. (69) admits four Killing vectors, namely, \begin{eqnarray} && \xi_0 = \left [ 1, 0, 0, 0 \right ] \: , \nonumber \\ && \xi_1 = \left [ - \frac{2 \alpha r_s \sin \phi}{\sin \theta}, 0, \cos \phi, - \sin \phi \cot \theta \right ] \: , \nonumber \\ && \xi_2 = \left [ - \frac{2 \alpha r_s \cos \phi}{\sin \theta}, 0, - \sin \phi, - \cos \phi \cot \theta \right ] \: , \nonumber \\ && \xi_3 = \left [ - 2 \alpha r_s, 0, 0, 1 \right ] \: , \end{eqnarray} for which, the following commutation relations hold \begin{eqnarray} && \left [ \xi_i \: , \: \xi_j \right ] = - \varepsilon_{ijk} \xi_k \: , \nonumber \\ && \left [ \xi_i \: , \: \xi_0 \right ] = 0 \: , \end{eqnarray} where $\varepsilon_{ijk}$ is the completely antisymmetric symbol of three indices, with $\varepsilon_{123} = 1 = \varepsilon_{r \theta \phi}$. 

At this point, we should stress that, the metric given by Eq. (69) allows for the same symmetries as the Taub-NUT metric~\cite{4},~\cite{5}, \begin{equation} ds^2 = \frac{R^2 - 2 \mu R - l^2}{R^2 + l^2} \left ( dT - 2 l \cos \theta d \phi \right )^2 - \frac{R^2 + l^2}{R^2 - 2 \mu R - l^2} dR^2 - \left ( R^2 + l^2 \right ) d \Omega^2 \: . \end{equation} In particular, 

\begin{itemize}

\item both of them have no curvature singularities;
\item provided that the values $\theta = 0$ and $\theta = \pi$ represent lines on the manifold, the time coordinate in both metrics is cyclic;
\item both metrics admit three Killing vectors, in addition to $\partial / \partial t$, and all of them obey the commutation relations given by Eqs. (71). 

\end{itemize}

In fact, the metrics given by Eqs. (69) and (72) represent the same line element, as it can be seen by performing in Eq. (72) the following transformation on the radial coordinate, \begin{equation} R = \left ( 1 + \alpha^2 \right )^{1/2} r - \frac{2 m \alpha^2}{\left ( 1 + \alpha^2 \right )^{1/2}} \: , \end{equation} together with a rescaling of the time coordinate in both Eqs. (69) and (72), \begin{equation} T = 2 l \tau \: , \qquad \qquad t = 4 \alpha m \tau \: , \end{equation} where \begin{equation} l = \frac{2 m \alpha}{\left ( 1 + \alpha^2 \right )^{1/2}} \: , \qquad \mu = \frac{m \left ( \alpha^2 - 1 \right )}{\left ( 1 + \alpha^2 \right )^{1/2}} \: . \end{equation} Notice that, in this case, there is an additional relation between the various parameters involved, namely, \begin{equation} \frac{\mu}{l} = \frac{\alpha^2 - 1}{2 \alpha} \: , \end{equation} i.e., a non-zero $\alpha$ controls the ratio $\mu/l$. Clearly, for $\alpha = \pm 1$, $\mu = 0$.

\section{Using the $\gamma$-metric as seed solution}

The $\gamma$-metric~\cite{7} generalizes the Schwarzschild solution and it can be used also as a seed metric to generate new solutions. In spherical coordinates, the $\gamma$-metric is given by \begin{eqnarray} ds^2 & = & \left ( 1 - \frac{2m}{r} \right )^{\gamma} dt^2 - \left ( 1 - \frac{2m}{r} \right )^{- \gamma} \left [ \left ( \frac{r^2 - 2 m r}{r^2 - 2 m r + m^2 \sin^2 \theta} \right)^{\gamma^2 - 1} dr^2 \right . \nonumber \\
& + & \left . \frac{\left ( r^2 - 2 m r \right)^{\gamma^2}}{\left ( r^2 - 2 m r + m^2 \sin^2 \theta \right )^{\gamma^2 - 1}} d \theta^2 + \left (r^2 - 2 m r \right ) \sin^2 \theta d \phi^2 \right ] \end{eqnarray} (see, e.g.,~\cite{6},~\cite{11}). The curved spacetime represented by the line element (77) possesses some very interesting features (see, e.g.,~\cite{12}). In particular, 

\begin{itemize}

\item if $\gamma \neq 0$ and $m = 0$, it corresponds to Minkowski spacetime; 
\item for $\gamma = 1$ and $m \neq 0$, it represents the Schwarzschild spacetime; 
\item for $\gamma \rightarrow \infty$ and $m \rightarrow 0$, but $m \gamma \rightarrow constant$, it represents a Curzon spacetime; 
\item the $\gamma$-solution is the static limit of the Tomimatsu-Sato (TS) family of solutions (see, e.g.,~\cite{1}). In this case, if the rotation parameter, $q$, vanishes, the TS solution becomes identical to the $\gamma$-solution, with the TS deformation parameter, $\delta$, being equal to the value of $\gamma$. 

\end{itemize}

The metric given by Eq. (77), is often referred to as the Voorhees metric or Zipoy-Voorhees (ZV) metric~\cite{7}. Upon consideration of the quaternionic version of the Ernst formalism, Hallilsoy~\cite{13} found a generalized version of the ZV metric. On the other hand, by treating the $\gamma$-solution as a seed (vacuum) metric, Richterek et al.~\cite{14},~\cite{15} obtained two new classes of solutions to the Einstein - Maxwell equations with interesting properties.

Here, on the basis of the technique of generating solutions presented in Sect. 3 (i.e., upon the action of a timelike Killing vector), we also come up with a new solution to the Einstein field equations in vacuum, which is written in the form \begin{eqnarray} ds^2 & = & \frac{\left ( r^2 - 2 m r \right )^{\gamma}} {r^{2 \gamma} + \alpha^2 (r - 2 m)^{2\gamma}} \left [ dt - 4 \alpha \gamma m \cos \theta d\phi \right ]^2 \nonumber \\ & - & \frac{r^{2 \gamma} + \alpha^2 (r-2m)^{2 \gamma}}{\left ( r^2 - 2 m \right )^{2\gamma}} \left [ \left ( \frac{r^2 - 2 m r}{r^2 - 2 m r + m^2 \sin^2 \theta} \right )^{\gamma^2 - 1} dr^2 \right . \nonumber \\ & + & \left . \frac{\left ( r^2 - 2 m r \right )^{\gamma^2}}{\left ( r^2 - 2 m r + m^2 \sin^2 \theta \right )^{\gamma^2 - 1}} d \theta^2 + \left ( r^2 - 2 m r \right ) \sin^2 \theta d \phi^2 \right ] \: . \end{eqnarray} A null tetrad (see, e.g.,~\cite{16}) and the Newmann-Penrose invariants (see, e.g.,~\cite{17}) of this new solution are given in the Appendix A. Accordingly, in view of Eqs. (A7) - (A12), we find that:

\vspace{.3cm}

In the case where $\alpha = 0$,

\begin{itemize}

\item for every $\gamma \neq 0, \: 1$, the Newmann-Penrose invariants $\Psi_i \; (i = 0, \: 1, ..., 4)$ coincide with those of the original $\gamma$-solution (as they should);
\item for $\gamma = 1$, only $\Psi_2$ is non zero, representing the corresponding quantity of the Schwarzschild solution.

\end{itemize}

In the case where $\alpha \neq 0$, we have $- \Psi_0 = \Psi_4 \neq 0$, $\Psi_1 = 0 = \Psi_3$ and $\Psi_2 \neq 0$. Consequently,

\begin{itemize}

\item for $\gamma \neq 0, \: 1$ and $\theta \neq 0, \: \pi$, the metric (78) is type I in Petrov's classification~\cite{18},~\cite{19}, exhibiting singularities at $r = 0$ and $r = 2m$. Singularities exist also at $r_{\pm} = m \left ( 1 \pm \cos \theta \right )$, if $\gamma \in (1, \: \sqrt{3})$;
\item for $\gamma = 1$ and $\theta \in (0,\pi)$, the metric is type D in Petrov's classification.

\end{itemize}

\section{Summary - Conclusions}

In this article, we have used a well-established method of generating stationary axisymmetric solutions to the Einstein field equations in vacuum (see, e.g., Chapter 34 of~\cite{1}), originating from already existing ones (seed solutions) that admit at least one Killing vector (either $\partial / \partial \phi$ or $\partial / \partial t$). 

Applying this method to the Schwarzschild solution, upon the action of the spacelike $\left ( \frac{\partial}{\partial \phi} \right )$ Killing vector, we have found a new family of vacuum solutions to the Einstein field equations, given by Eq. (38) (or Eq. 41). In the same reasoning, but this time upon the action of the timelike $\left ( \frac{\partial}{\partial t} \right )$ Killing vector, the Schwarzschild solution generates the family of solutions given by Eq. (58) (or Eq. 69), which, upon the transformation given by Eqs. (73) - (75), coincides to the Taub-NUT metric, given by Eq. (72). Finally, using as seed solution the $\gamma$-metric (77), and upon the action of the timelike Killing vector, we have arrived at a new family of solutions to the Einstein field equations, given by Eq. (78), which, for $\gamma = 1$ and $\theta \in (0, \pi)$, represents an asymptotically flat spacetime.

Notice that, upon consideration of this particular method of generating solutions to the Einstein field equations in vacuum, a free parameter, $\alpha$, is introduced, which, throughout our analysis, was taken to be real. As recognized by Harrison~\cite{20}, in the case where this parameter is complex it mixes gravity with electromagnetism, since the quantity $H = \frac{1}{2} \alpha \alpha^{\star}$ represents the strength of a uniform magnetic field. On the other hand, if $\alpha$ is purely imaginary, it represents the potential of the electromagnetic field and there is no twist.

\vspace{.5cm}

\textbf{Acknowledgments:} This work has been supported by the General Secretariat for Research \& Technology of Greece and by the European Social Fund, within the framework of the action "EXCELLENCE". 

\section*{Appendix A}

In this Appendix, we present the Newmann-Penrose scalars~\cite{17} of the metric given by Eq. (78), which is generated from the $\gamma$-metric (77). In the metric given by Eq. (78), we set $$ f = \frac{\epsilon}{1 + \alpha^2 \epsilon^2} = \frac{(r^2 - 2 m r)^{\gamma}}{r^{2 \gamma} + \alpha^2 (r - 2m)^{2 \gamma}} \eqno{(A1)} $$ Accordingly, we consider the following set of orthonormal vectors (see, e.g.,~\cite{16}) $$ A_{\mu} = \left [ \sqrt{f}, 0, 0, - 4 \alpha m \cos \theta \sqrt{f} \right ] $$ $$ B_{\mu} = \left [ 0, - \frac{1}{\sqrt{f}} \left (\frac{r^2 -2 m r}{r^2 - 2 m r + m^2 \sin^2 \theta} \right )^{\frac{\gamma^2 - 1}{2}}, 0, 0 \right ] $$ $$ P_{\mu} = \left [ 0, 0, \frac{1}{\sqrt{f}} \frac{\left ( r^2 - 2 m r \right )^{\frac{\gamma^2}{2}}}{\left ( r^2 - 2mr + m^2 \sin^2 \theta  \right )^{\frac{\gamma^2-1}{2}}}, 0 \right ] $$ $$ Q_{\mu} = \left [ 0, 0, 0, \frac{\sin \theta}{\sqrt{f}} \sqrt{r^2 - 2mr} \right ] \: , \eqno{(A2)} $$ from which, a null tetrad for the metric (78) can be determined, as follows
$$ l_{\mu} = \frac{B_{\mu} + A_{\mu}}{\sqrt{2}} \: ,~~k_{\mu} = \frac{B_{\mu} - A_{\mu}}{\sqrt{2}} $$ $$ m_{\mu} = \frac{P_{\mu} + i Q_{\mu}}{\sqrt{2}} \: ,~~\bar{m}_{\mu} = \frac{P_{\mu} - i Q_{\mu}}{\sqrt{2}} \eqno{(A3)} $$ resulting in $$ l_{\mu} = \frac{1}{\sqrt{2}} \left [ \sqrt{f}, - \frac{1}{\sqrt{f}} \left (\frac{r^2 - 2mr}{r^2 - 2mr + m^2 \sin^2 \theta} \right )^{\frac{\gamma^2-1}{2}}, 0, - 4 \alpha m \cos \theta \sqrt{f} \right ] $$ $$ l^{\mu} = \frac{1}{\sqrt{2}} \left [ - \frac{1}{\sqrt{f}} \left ( \frac{r^2 - 2mr + m^2 \sin^2 \theta}{r^2 - 2mr} \right )^{\frac{\gamma^2-1}{2}}, - \sqrt{f}, 0, 0 \right ] $$ $$ k_{\mu} = \frac{1}{\sqrt{2}} \left [ - \sqrt{f}, - \frac{1}{\sqrt{f}} \left ( \frac{r^2 - 2mr}{r^2 - 2mr + m^2 \sin^2 \theta} \right )^{\frac{\gamma^2-1}{2}}, 0, 4 \alpha m \cos \theta \sqrt{f} \right ] \eqno{(A4)} $$ $$ k^{\mu} = \frac{1}{\sqrt{2}} \left [ \frac{1}{\sqrt{f}}, - \sqrt{f} \left ( \frac{r^2 - 2mr + m^2 \sin^2 \theta}{r^2 - 2mr} \right )^{\frac{\gamma^2-1}{2}}, 0, 0 \right ] $$ $$ m_{\mu} = \frac{1}{\sqrt{2}} \left [ 0, 0, \frac{1}{\sqrt{f}} \frac{\left ( r^2 - 2mr \right )^{\frac{\gamma^2}{2}}}{\left ( r^2 - 2mr + m^2 \sin^2 \theta \right )^{\frac{\gamma^2-1}{2}}}, \frac{i \sin \theta}{\sqrt{f}} \sqrt{r^2 - 2mr} \right ] $$ $$ m^{\mu} = \frac{\sqrt{f}}{\sqrt{2} \sqrt{r^2 - 2mr}} \left [ \frac{4i \alpha m \cos \theta}{\sin{\theta}}, 0, \left ( \frac{r^2 - 2mr + m^2 \sin^2 \theta}{r^2 - 2mr} \right )^{\frac{\gamma^2-1}{2}}, - \frac{i}{\sin \theta} \right ] \: . $$ The tetrad given by Eqs. (A4) satisfies the orthonormality conditions $$ l^{\mu} l_{\mu} = k^{\mu} k_{\mu} = m^{\mu} m_{\mu} = 0 $$ $$ l^{\mu} m_{\mu} = k^{\mu} m_{\mu} = 0 $$ $$ l^{\mu} k_{\mu} = 1 \: ,~~m^{\mu} \bar{m}_{\mu} = 1 \: . \eqno{(A5)} $$ Furthermore, in view of Eqs. (A4), we verify that the components of the metric tensor given by Eq. (78) may obtained from the relation~\cite{16} $$ g_{\mu \nu} = l_{\mu} k_{\nu} + l_{\nu} k_{\mu} + m_{\mu} \bar{m}_{\nu} + m_{\nu} \bar{m}_{\mu} \: . \eqno{(A6)}$$ 

Now, following~\cite{17}, the Newmann-Penrose invariants corresponding to the solution (78), that is generated from the $\gamma$-metric by the action of the $t$-Killing vector, are given by $$ \Psi_0 = 2 \: {\cal C}_{\mu \nu \kappa \lambda} l^{\mu} l^{\nu} m^{\kappa} m^{\lambda} \: \Rightarrow $$ $$ \Psi_0 = - \frac{\gamma(\gamma^2-1)m^3(r-m)}{2 \left [ r^{2\gamma}+\alpha^2(r-2m)^{2\gamma} \right ]^2} \times \frac{\left ( r^2 - 2mr + m^2 \sin^2 \theta \right )^{\gamma^2-2}}{(r^2-2mr)^{\gamma^2-\gamma+1}} $$ $$ \times \left \lbrace \left [ r^{2\gamma}-\alpha^2(r-2mr)^{2\gamma} \right ] + 2 \imath \left ( r^2-2mr \right )^{\gamma} \right \rbrace \: , \eqno{(A7)} $$ where ${\cal C}_{\mu \nu \kappa \lambda}$ is the Weyl tensor. Regarding the rest of the Newmann-Penrose invariants, we have $$ \Psi_1 = -{\cal C}_{\mu \nu \kappa \lambda} l^{\mu} m^{\nu} \left ( k^{\kappa} l^{\lambda} + \bar{m}^{\kappa} m^{\lambda} \right ) = 0 \eqno{(A8)} $$ and \begin{eqnarray} \Psi_2 & = & 2 \: C_{\mu \nu \kappa \lambda} l^{\mu} m^{\nu} n^{\kappa} \bar{m}^{\lambda} \nonumber \\ &&=-\frac{Z^{\gamma^2-3}}{8\sin^2{\theta} Y^5 X^{\gamma^2-\gamma+2}} \left \lbrace -4m^2(\gamma^2-1)Z Y^4 X^2\sin^2{\theta}\cos{2\theta} \right . \nonumber\\
&&-4(\gamma^2-1)Z Y^4 X^3\sin^2{\theta}\nonumber\\
&&+8\gamma(2\gamma-1)Z^2 Y X^2(XY^2\sin^2{\theta}+8\alpha^2 m^2\gamma^2\cos^2{\theta} X^{2\gamma}Y)\nonumber\\
&&+4 Z^2 Y^2 X[(\gamma-1)^2 Y^2\sin^2{\theta}+16\alpha^2 m^2\gamma^3\cos^3{\theta} X^{2\gamma}]\nonumber\\
&&+8(\gamma^2-1)m^4 Y^4 X\sin^4{\theta}\cos^2{\theta}\nonumber\\
&&-4(\gamma^2-1)m^2Z Y^2 X\sin^2{\theta}\cos^2{\theta}[XY^2+16\alpha^2 m^2\gamma^2 X^{2\gamma}]\nonumber\\
&&-8\gamma^2 Z^2 X^2[Y^2 \left ( 2X+X^{-\gamma^2+\gamma+1} \right ) \sin^2{\theta}+32\alpha^2m^2\gamma^2\cos^2{\theta}X^{2\gamma}][r^{2\gamma-1}+\alpha^2(r-2m)^{2\gamma-1}]^2\nonumber\\
&&-4\gamma(r-m)Z^2 Y^3 X\sin^2{\theta}[(\gamma^2-3) X+(\gamma^2-4\gamma+1) X^{-\gamma^2+\gamma+1}][r^{2\gamma-1}+\alpha^2(r-2m)^{2\gamma-1}]\nonumber\\
&&+4\gamma(\gamma^2-1)(r-m)Z Y X^2[Y^2[X+X^{-\gamma^2+\gamma+1}]\sin^2{\theta}\nonumber\\
&&+16\alpha^2m^2\gamma^2\cos^2{\theta} X^{2\gamma}][r^{2\gamma-1}+\alpha^2(r-2m)^{2\gamma-1}]\nonumber\\
&&-4\gamma(\gamma-1)(r-m)^2 ZY^2X[Y^2[X+X^{-\gamma^2+\gamma+1}]\sin^2{\theta}+16\alpha^2m^2\gamma^2\cos^2{\theta} X^{2\gamma}]\nonumber\\
&&+4\gamma(\gamma-1)^2(r-m)^2Z^2Y^2[Y^2[X+X^{-\gamma^2+\gamma+1}]\sin^2{\theta}+16\alpha^2m^2\gamma^2\cos^2{\theta} X^{2\gamma}] \nonumber \end{eqnarray} $$ \left . ~~~~~+ \: 8(\gamma^2-1)(r-m)^2Y^4X^3\sin^2{\theta} + 32\alpha^2m^2\gamma^2Z^2Y^2X^{-\gamma^2+3\gamma+1}\sin^2{\theta} \right \rbrace \: , \eqno{(A9)} $$ where we have set $$ X = r^2 - 2mr \: ,~~Y = r^{2\gamma} + \alpha^2(r-2m)^{2\gamma} \: ,~~\mbox{~and~}~~Z = r^2 - 2mr + m^2 \sin^2 \theta \: . \eqno{(A10)} $$ Finally, $$ \Psi_3 = - \: {\cal C}_{\mu \nu \kappa \lambda} k^{\mu} \bar{m}^{\nu} \left ( k^{\kappa} l^{\lambda} + \bar{m}^{\kappa} m^{\lambda} \right ) = 0 \eqno{(A11)} $$ and $$ \Psi_4 = 2 \: {\cal C}_{\mu \nu \kappa \lambda} n^{\mu} n^{\nu} \bar{m}^{\kappa} \bar{m}^{\lambda} = - \Psi_0 \: . \eqno{(A12)} $$

\end{document}